# SoniMet – A tool for sonifying and visualizing the performance of single researchers

Tim Waterfield$, Lutz Bornmann*

$Website: https://karhide.co.uk/

Email: info@kennebec.co.uk

*Science Policy and Strategy Department

Administrative Headquarters of the Max Planck Society

Hofgartenstr. 8,

80539 Munich, Germany.

E-Mail: bornmann@gv.mpg.de

**Abstract**

For centuries, the scientific community has predominantly relied on visual tools to communicate empirical results and complex datasets. While visual representations dominate bibliometric analyses, the human auditory system possesses sensitive capacities for processing complex temporal information, distinguishing intricate patterns, and tracking parallel data streams. Data sonification translates data relations into acoustic signals, offering an alternative method for data exploration, pattern recognition, and scientific communication. This paper applies this concept to the field of bibliometrics through metrics sonification—the auditory translation of bibliometric information—and introduces SoniMet (Sonifying Metrics), a web-based tool designed to visualize and sonify the publication and citation data of individual scientists (see https://sonimet.kennebec.co.uk). SoniMet connects to the OpenAlex database to retrieve bibliometric records and displays them on an interactive chronological timeline. The tool employs direct parameter mapping to translate citation impact indicators into non-speech sound: field-weighted citation impact and citation counts determine the pitch and volume of a synthesized note and are mapped to the acoustic echo strength. Although SoniMet expands the methodological toolkit for research evaluation, current limitations include its restriction to individual scholar profiles and the challenge of integrating transient audio files into traditional, text-based scientific publishing workflows. Future empirical user studies are necessary to systematically evaluate the analytical utility and cognitive benefits of metrics sonification compared to established visual methods.

# 1 Introduction

For centuries, the scientific community has relied on visual tools to communicate empirical results and complex datasets. The visualization of results based on publication and citation data is also prevalent in bibliometrics. Western culture operates primarily as a visual culture, relying on the perceptual capacities of the eye. However, the human body possesses other sensitive sensory organs. The auditory system is adept at processing complex temporal information and distinguishing intricate patterns. Given these biological capabilities, researchers across various disciplines have begun exploring auditory display techniques to complement traditional data visualization, recognizing that the use of multiple human senses can lead to a profound understanding of information (Dayé & de Campo, 2006).

The practice of using sound for scientific observation has historical precedents. For example, Galileo Galilei utilized the rhythmic sounds of a rolling ball striking catgut strings to analyze physical motion. In contemporary research, data sonification is formally defined as the transformation of data relations into perceived relations in an acoustic signal to facilitate communication or interpretation (Kramer et al., 2010). An extended definition describes it as a systematic, describable, and reproducible technique that translates data into non-speech sound to reveal latent meaning, serving practical, artistic, or scientific purposes (Liew & Lindborg, 2020). Depending on the specific application, auditory displays can be broadly categorized into alarms, status monitoring messages, and exploratory data analysis tools (Walker & Nees, 2011). Various methods exist within the field of sonification. Audification is the most direct approach, translating data waveforms straight into the audible domain, which is particularly effective for large time-series datasets (Dombois & Eckel, 2011).

Parameter mapping has emerged as the most widely used technique in data sonification (Dubus & Bresin, 2013). In parameter mapping, specific data dimensions are systematically mapped to auditory parameters such as pitch, volume, duration, and timbre (Grond & Berger, 2011). Pitch is the most frequently used auditory dimension because the human ear is sensitive to minor frequency changes, and musical scales offer pre-existing cognitive structures that facilitate data interpretation (Neuhoff, 2011). Other approaches include auditory icons. They use brief environmental sounds metaphorically linked to data events, and earcons, which use structured sequences of synthetic tones (Walker & Nees, 2011).

Data sonification offers several advantages over visual displays. First, the human auditory system excels at recognizing temporal patterns, periodic events, and subtle transient changes in continuous data streams (Kramer et al., 2010; Zanella et al., 2022). Sound is inherently multi-dimensional, allowing users to listen to different sonified streams in parallel and detect weak signals amidst background noise, a phenomenon akin to the cocktail party effect (Dombois & Eckel, 2011; Zanella et al., 2022). Second, hearing is a continuous and omnidirectional sense, which makes sonification suitable for monitoring tasks in environments where visual attention is restricted or occupied, such as in medical operating theaters or aviation (Walker & Nees, 2011). Third, sonification enhances accessibility, enabling visually impaired researchers and students to analyze complex datasets and participate equally in scientific analysis (Zanella et al., 2022). Fourth, beyond practical utility, sonification has a unique capacity to trigger emotional responses and aesthetic engagement, which can be effective in public communication and science outreach (Lenzi & Ciuccarelli, 2020).

Despite its potential, sonification faces significant challenges. First, unlike visual graphs, which are systematically taught and universally understood from a young age, identifying structures in a given sequence of sounds requires specialized training and active listening skills (Dayé & de Campo, 2006). Second, there is a risk that aesthetic and musical intentions might overshadow the objective communication of data, limiting the scientific utility of sonification (Vickers, 2017). This tension forms part of the broader mapping problem, which underscores the subjective nature of assigning data values to acoustic dimensions (Lenzi & Ciuccarelli, 2020; Supper, 2015). For instance, mapping data to a traditional Western musical scale inherently imposes cultural biases on the interpretation (Tulilaulu et al., 2018). Third, the sonification field lacks universally accepted standardized methods, complicating reproducibility and broader mainstream adoption (Zanella et al., 2022). To overcome these hurdles, successful sonification projects require interdisciplinary collaboration between domain scientists, audio engineers, and sound designers, ensuring that technical execution does not obscure empirical interpretation (Supper, 2015).

Successful applications of sonification date back to early measurement devices like the pulse oximeter, which conveys blood oxygen levels through pitch variations during surgery (Kramer et al., 2010). More recently, numerous complex datasets have been transformed into auditory experiences, many of which are cataloged in the Data Sonification Archive (Lindborg et al., 2024). The Microbial Bebop project, for example, mapped oceanic microbial activity and environmental parameters into jazz compositions, demonstrating the potential of sonification for public engagement (Larsen & Gilbert, 2013). In the context of climate change and ecology, the Loud Numbers project sonified the decline of Danish insect populations over two decades; this project utilized pitch to represent insect size and a descending melody to signify the ongoing loss of biodiversity (Lindborg et al., 2023). Another notable project is Bristol Burning (see https://youtu.be/7QxkM7cLOuI?si=RDunavg9FxIXSP3C), which merged air quality data with a hip-hop track to educate the public on pollution metrics through an immersive experience. In astronomy, sonification has been used to detect gravitational waves and classify galaxy spectra, combining sound with visual components to maximize analytical efficacy (Zanella et al., 2022).

Building on these developments in the natural and social sciences, data sonification can be also adapted to the field of bibliometrics. Bornmann and Haegner (2025) define this specialized application as metrics sonification: the auditory translation of bibliometric

information—such as publication output and citation impact—into sound for analytical and communicative purposes. Social science data, including bibliometric datasets, are particularly well-suited for sonification because they depict complex, multidimensional relations and temporal interdependencies (Dayé & de Campo, 2006). One of the authors (LB) has previously explored the potential of metrics sonification in several projects.

In one study, Bornmann and Haegner (2025) sonified the publication data of Loet Leydesdorff, a seminal figure in scientometrics, to honor his academic legacy and communicate his career trajectory. By mapping properties such as publication output, open access status, and field-normalized citation impact to pitch variations in an F minor scale, Bornmann and Haegner (2025) combined quantitative parameter mapping with spoken audio to provide essential context. In another project, Bornmann and Leibel (2025) investigated the specification uncertainty in bibliometric indicators by sonifying the effect sizes of large versus small research teams on disruptive research. Using results from a multiverse analysis comprising several regression models from an empirical project conducted by Leibel and Bornmann (2026), divergent effect sizes were mapped to pitch in C major to audibly demonstrate methodological variability. Bornman and Ganser (2025) compared the temporal publication output of hyperprolific academic authors with average scientists by assigning distinct audio signatures and temporal frequencies to publication events, illustrating the vast differences in scholarly productivity in a visceral manner.

There are already numerous tools in the scientometric community, such as VOSviewer (van Eck & Waltman, 2010) or CiteSpace (Chen, 2006), to visualize bibliometric data. With SoniMet (Sonifying Metrics), we now introduce a tool to not only visualize but also sonify publication and citation data of individual scientists. SoniMet offers an interactive publication timeline where specific metrics, such as the field-weighted citation impact (FWCI), are mapped to pitch and volume, while the citation count determines the echo strength. By enabling users to explore a scholar’s career analytics through both visual elements and custom synthesized notes, SoniMet exemplifies the synergy between scientometrics and auditory design, expanding the methodological toolkit for research evaluation and communication.

# 2 How does SoniMet work?

SoniMet is a web-based tool designed for the interactive, visual, and auditory presentation of publication and citation data (see https://sonimet.kennebec.co.uk). At https://sonimet.kennebec.co.uk/guide, the user finds a guide to the tool that describes all its features in detail.

To retrieve the data, the tool utilizes an application programming interface (API) to connect to OpenAlex (Priem et al., 2022). OpenAlex is a comprehensive, open-source bibliographic database and catalog of global scholarly publications, authors, and institutions. Through this API connection, SoniMet resolves scholar profiles to access their respective bibliometric records. When a user enters a name in the search field (first and last name), the system typically displays various author names along with their affiliated institutions (which were found in OpenAlex under the searched name). The user can then select the names that correspond to the scholar they are looking for. Publication metadata for the selected names is retrieved from OpenAlex (including the h index, total citations and number of works that are displayed alongside the name in SoniMet).

The central visual component of the tool is the 'Scholar Timeline' (see Figure 1). In this chronological display, an author's individual publications are represented as data points, with the system visually distinguishing between Open Access (OA) and Closed Access (CA) articles. This timeline is fully interactive: hovering over or clicking on a publication dot plays a custom synthesized note specific to that work and simultaneously opens a 'Selected Work Details' panel. This panel displays comprehensive metadata for the selected publication, including the publication year, work title, first author, authors count, source journal or venue, and a DOI link to the publication. The sonification for an author can be played not only at different tempos (‘Tempo’), but also in different orders of the indicator values (‘Order’, see Figure **1**).

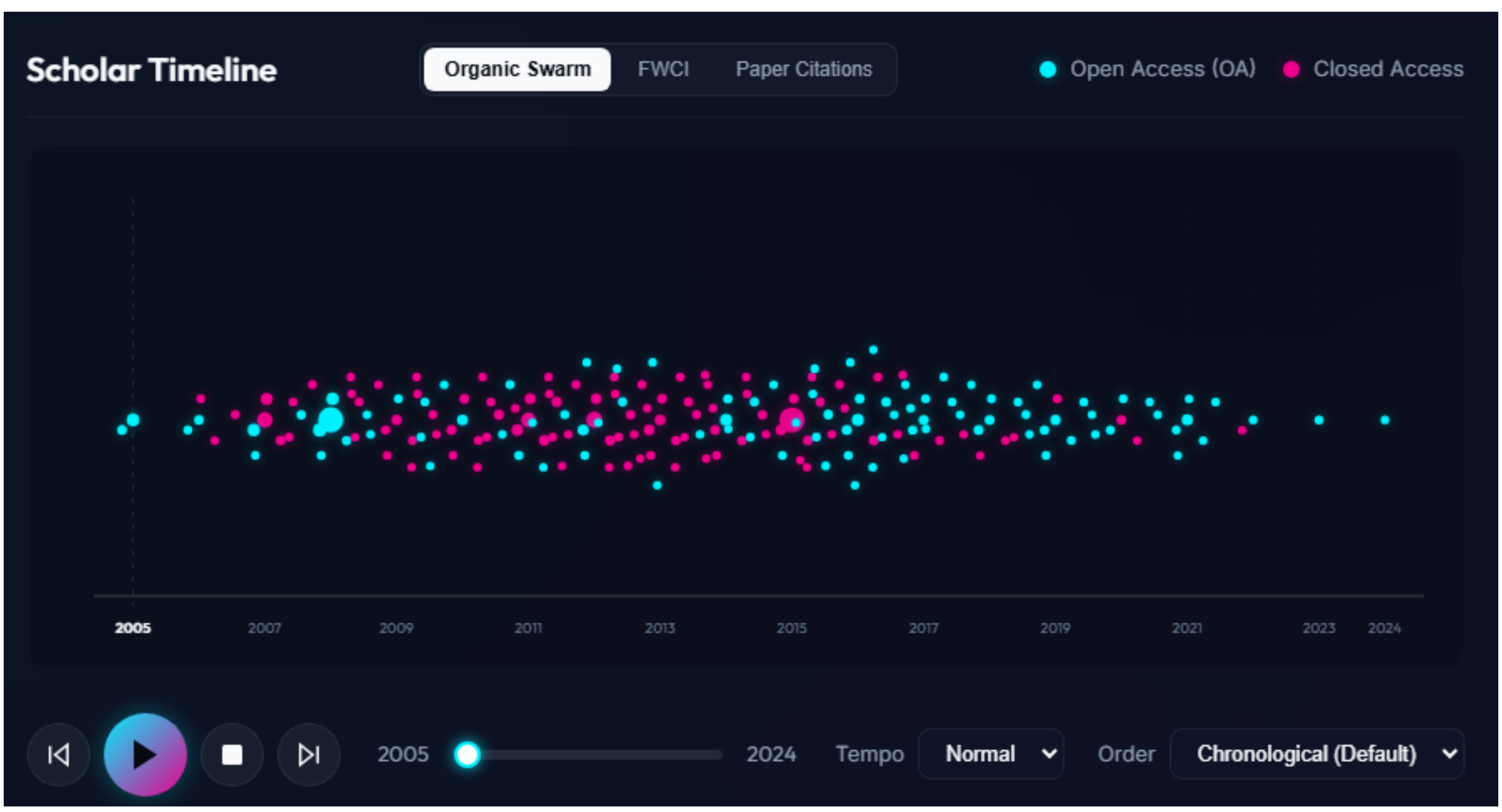

Figure 1. ‘Scholar Timeline’ in SoniMet

The sonification of the data in SoniMet is achieved through direct parameter mapping, which translates specific citation impact indicators into acoustic signals: citation count and FWCI. The FWCI is a metric that measures a publication’s citation performance by comparing its actual number of citations against the global average expected for publications of the same subject field and publication year. In SoniMet, FWCI and citation count are mapped directly to the pitch and volume of the audio signal. For example, a citation impact of 1.8 is represented by the pitch D4 at a frequency of 293.7 Hz. Volume corresponds directly to citation volume: highly cited papers sound loud and prominent, while papers with few citations sound soft. The citation impact is additionally mapped to the echo feedback strength. For instance, a publication with 45 citations generates a long, lingering echo of 85% feedback strength, creating an acoustic ripple that represents its sustained academic legacy. Co-authorship size is sonified through dynamic, synchronized percussion, translating the collaboration scale of each work into a rolling drum pattern.

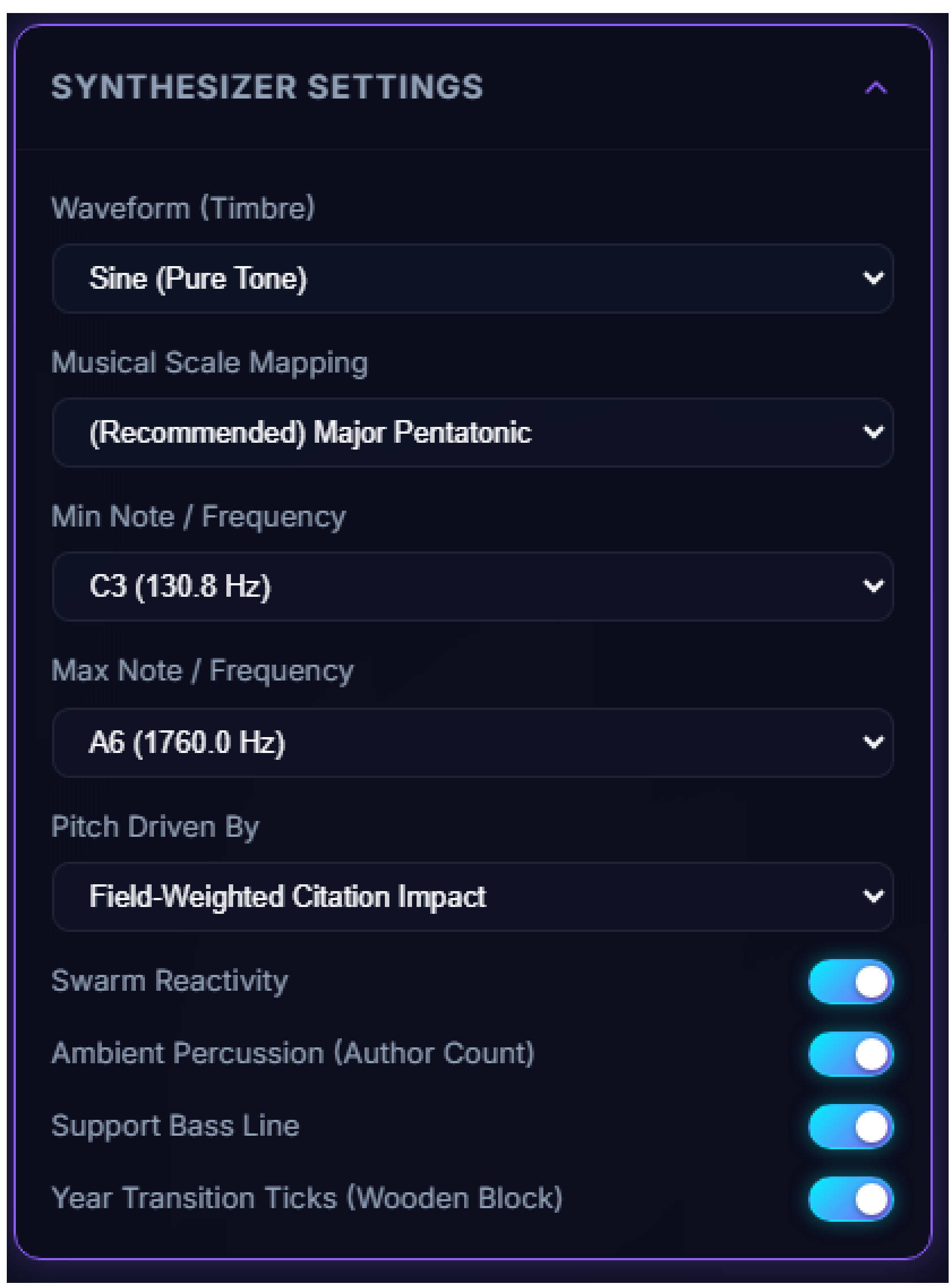

Figure 2. ‘Synthesizer Settings’ in SoniMet

The tool’s auditory parameters are highly customizable via the ‘Synthesizer Settings’ panel (see Figure 2). Users can select the waveform—ranging from pure sine waves to retro square-wave timbres—which determines the fundamental texture of the chime. The musical scale mapping allows for the quantization of data into specific intervals, such as Major or Minor Pentatonic scales for harmonic coherence, or a continuous frequency scale for analog-style pitch glides. The pitch range is constrained by adjustable Min/Max Note or Frequency limits, ensuring sonification remains within a comfortable perceptual register. Users can designate which OpenAlex metadata field—citation count, FWCI, or a combined impact index—drives the pitch modulation. Additional parameters provide interactive and contextual cues: the ‘Swarm Reactivity’ toggle synchronizes visual node expansion with audio playback, ‘Ambient Percussion’ generates a rhythmic drum track linked to co-author counts, ‘Support Bass Line’ adds a low-frequency harmonic foundation at year boundaries, and ‘Year Transition Ticks’ provide an acoustic marker for the temporal (annual) progression of the timeline.

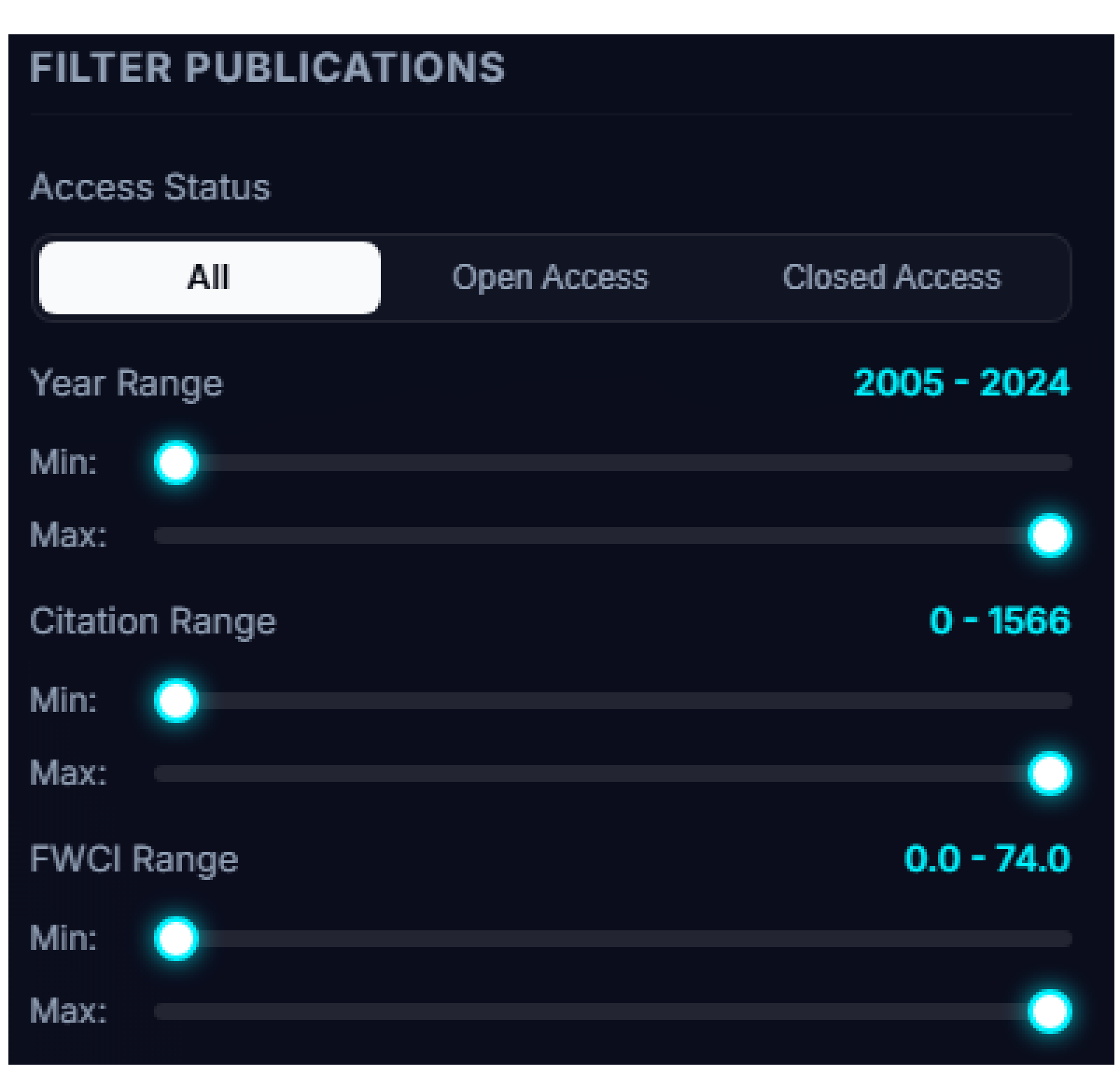

Figure 3. ‘Filter Publications’ in SoniMet

In addition to visualization and sonification, the tool includes filter mechanisms (see Figure 3) and analytical functions for data exploration (see Figure 4). Users can restrict the displayed publications by defining minimum and maximum values across several parameters, including publication year (e.g., 2004 to 2023), citation count (e.g., 0 to 100), and FWCI (e.g., 0.0 to 1.0). In the 'Scholar Career Analytics' section, SoniMet presents the overall performance of the researcher analyzed (retrieved from OpenAlex), including number of works, the OA ratio, the average FWCI, and the average citations per work.

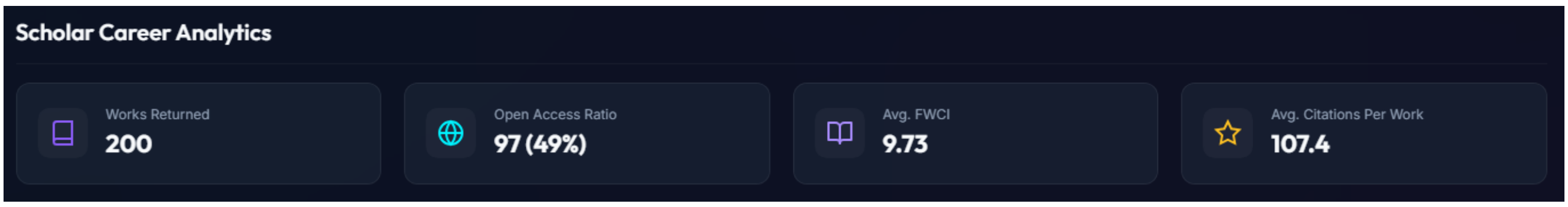

Figure 4. 'Scholar Career Analytics' in SoniMet

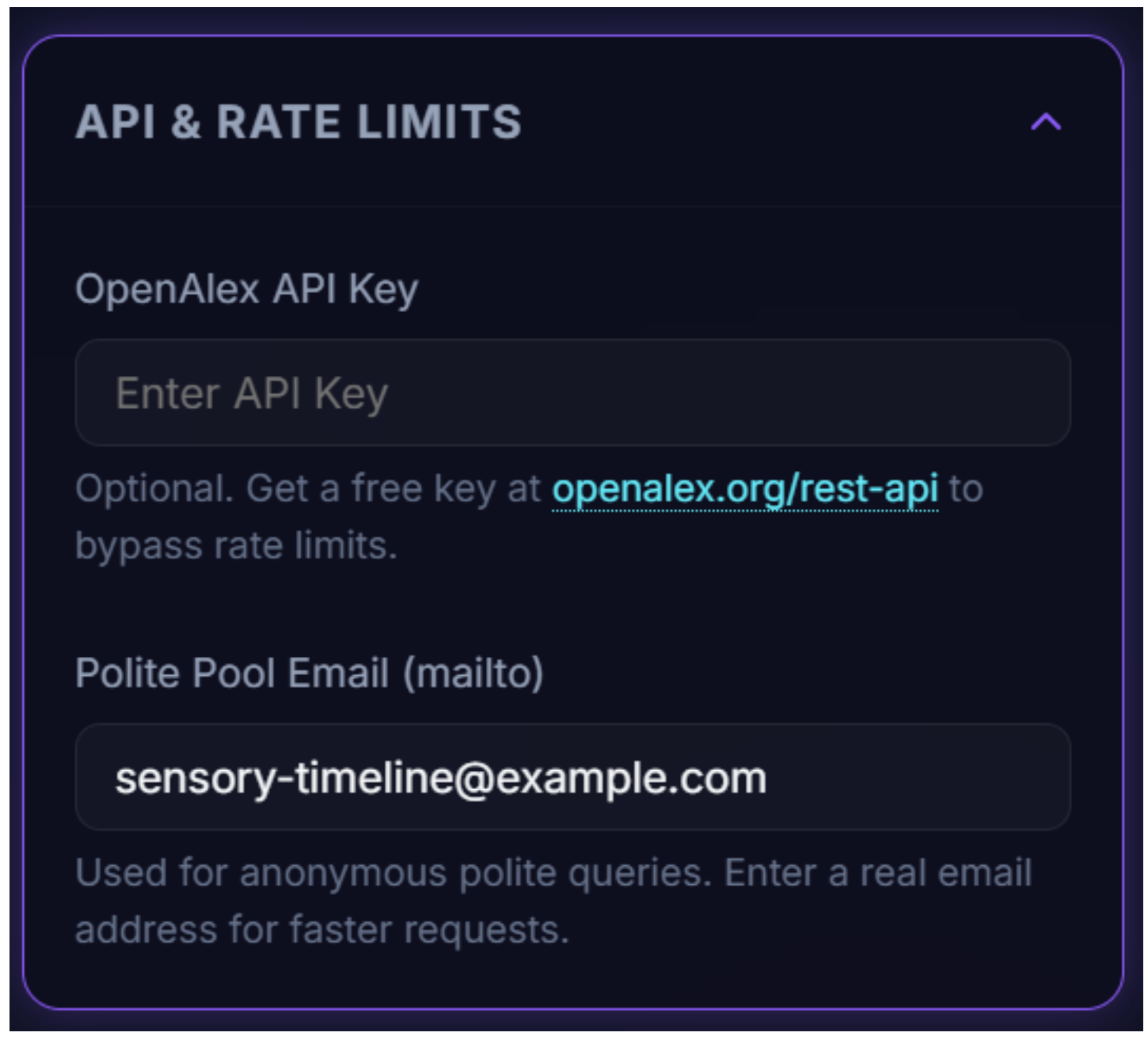

Figure 5. 'API & Rate Limits' in SoniMet

To mitigate potential search query failures caused by high traffic volume on the public OpenAlex API, SoniMet allows users to customize their connection parameters within the 'API & Rate Limits' settings panel (see Figure 5). Users can provide a valid email address to access the OpenAlex 'Polite Pool' for prioritized response times, or enter a personal, free OpenAlex API Key to bypass standard anonymous rate limits. To enhance retrieval efficiency and avoid text search constraints, users may input a direct OpenAlex Author ID or the complete author URL into the search field, enabling the tool to fetch data directly.

For documentation, presentation, and validation purposes, SoniMet offers export functionalities, allowing users to download the generated audio (‘Export Sonification’), the visual representation (‘Export Timeline Graph’), and the visualized and sonified data from OpenAlex as a CSV file (see Figure 6). Visual representation can be included in text documents to present the output and citation impact performance of single researchers. A look at a researcher’s career trajectory illustrates how publication output, citation impact, and the ratio of OA to CA have changed. The audio generated contains the same information; however, it appeals to a different sensory channel: the ear. The audio version of the data can be downloaded as a WAV file or in MIDI format. The MIDI format can be used to further process the track in a digital audio workstation (DAW), such as Ableton Live (see https://www.ableton.com). For example, one author (TW) used the MIDI files from publication and citation data of Nobel laureates 2025 to produce a music album. The album can be listened to on SoundCloud (see https://soundcloud.com/karhide/sets/sonimet). We would appreciate it if users would cite the tool when they use it in their own music production.

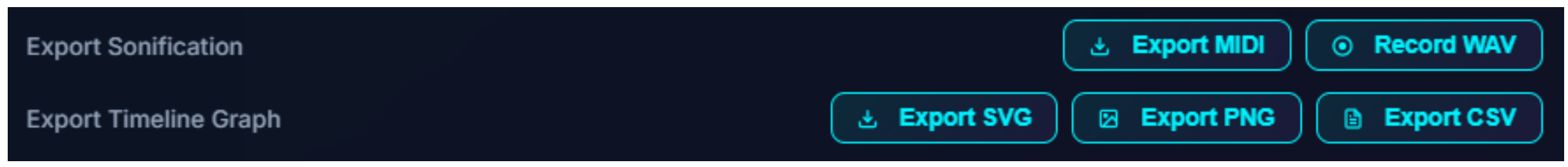

Figure 6. Export options in SoniMet

Since the publications associated with author IDs in OpenAlex are not always 100% correct (i.e. publications have not been correctly assigned to a distinct person), the user can download the underlying publication and citation impact data used for the sonification and visualization (‘Export CSV, see Figure 6). The user can use the export file to check whether the publications for an author are complete or incomplete.

# 3 Discussion

The visual representation of empirical findings has long been the dominant method for communication within the scientific community, and bibliometrics is no exception. Scientific culture operates primarily as a visual culture, relying heavily on the perceptual capacities of the eye to read graphical displays, scatter plots, and science maps (Dayé & de Campo, 2006). However, the human sensory system offers other sensitive capacities that remain underutilized in empirical research. The auditory system, in particular, is adept at processing complex temporal information, distinguishing intricate patterns, and tracking multiple streams of simultaneous events (Kramer et al., 2010). Recognizing these biological capabilities, data sonification has emerged as a method that translates data relations into perceived relations in an acoustic signal, aiming to facilitate communication, interpretation, and the discovery of latent meaning (Liew & Lindborg, 2020). By adapting these techniques to bibliometric information, metrics sonification provides an alternative method to present publication outputs, citation impacts, and structural dynamics through sound (Bornmann & Haegner, 2025).

With SoniMet, we have introduced a web-based tool designed to complement traditional visual bibliometrics by offering an interactive, visual, and auditory presentation of publication and citation data for individual scientists. The tool connects to the OpenAlex database via an application programming interface to retrieve bibliometric records. It presents the data on an interactive ‘Scholar Timeline’ that visually distinguishes between OA and CA publications. SoniMet employs parameter mapping, the most widely used sonification technique (Dubus & Bresin, 2013; Grond & Berger, 2011), to translate bibliometric indicators into acoustic signals. Specifically, the FWCI and citation count of a publication are mapped to the pitch and volume of a synthesized tone and mapped to the strength of an acoustic echo. Through interactive filtering and aggregated career analytics, SoniMet allows users to explore a researcher’s output both visually and acoustically.

Despite the conceptual advantages of adding an auditory dimension to bibliometric analysis, the current version of SoniMet presents several practical and methodological limitations.

A first limitation is the tool’s restriction to the analysis of individual scholar profiles from OpenAlex. At present, users cannot compare multiple researcher profiles. Since bibliometric evaluations frequently require comparative analyses between groups of researchers, departments, or entire research fields (Lepori & Bornmann, 2026), the inability to sonify comparative datasets restricts the tool’s utility for broader evaluative contexts.

A second limitation concerns the integration of the exported sonification into traditional scientific workflows. SoniMet allows users to export both the visual timeline graph and the generated audio files. While a visual graph can be seamlessly inserted into a printed manuscript, a sonification cannot be embedded in traditional printed papers (Dayé & de Campo, 2006). Although digital publishing formats increasingly support multimedia supplements, it remains unclear how researchers can effectively utilize exported audio files in scientific communication, which relies heavily on static, visual text documents.

The third limitation concerns an open question regarding whether users will find metrics sonification as helpful as traditional visualization. Decades of formal education have trained scientists to interpret visual graphs, scatter plots, and network maps with a high degree of sophistication (Dayé & de Campo, 2006). In contrast, identifying quantitative structures in a sequence of sounds requires specialized training, active auditory attention, and a process referred to as ‘reduced listening’ (Zanella et al., 2022). Without standardized mapping conventions and appropriate listener training, sonification might appear arbitrary to untrained users, limiting its immediate analytical usefulness compared to well-established visual alternatives.

SoniMet integrates audio elements, such as bass, percussions or rhythmic structures, to transform the sonification into a more cohesive musical experience. Research in auditory displays has shown that incorporating musical structures can reduce listener fatigue and increase aesthetic engagement (Vickers, 2017). By carefully introducing rhythmic elements, data points could be contextualized within a temporal grid, making chronological publication patterns more discernible. However, the integration of musical elements should be approached cautiously why we would like to mention it as fourth limitation. There is an inherent tension in sonification design between objective data representation and subjective aesthetic choices (Vickers, 2017). If aesthetic intentions overshadow the objective communication of the underlying data, the scientific utility of the sonification is compromised. Therefore, the addition of percussive and musical elements should remain strictly data-driven, ensuring that the sonification maintains its analytical integrity while becoming a more pleasant auditory experience.

In future research and development, we plan to address the limitations of SoniMet by expanding its functional scope. For example, we could expand on proceeding multiple authors and how they interact. One thought would be different instrumentation for each of the individuals so the user could build a band or orchestra around a research group. Another extension could be to enable the visualization and sonification of aggregated entities, such as the performance of specific academic journals, research institutions, or entire countries. Sonifying larger, more complex datasets could leverage the auditory system’s ability to detect subtle transient changes and identify patterns in noisy data environments, a phenomenon related to the cocktail party effect (Kramer et al., 2010; Zanella et al., 2022).

To ascertain the true value of metrics sonification, it is imperative to evaluate SoniMet in future empirical user studies. Systematic evaluation is necessary to determine whether sonification genuinely assists users in understanding bibliometric data, or if it merely serves as an engaging novelty. Future studies should assess user comprehension, cognitive load, and modality preferences when interacting with SoniMet's audio-only, visual-only, and combined audio-visual interfaces. Past research on sonification projects provides a strong foundation for such evaluations and indicates that auditory displays can yield measurable benefits. For example, in the medical field, an evaluation conducted at the Georgia Institute of Technology's Sonification Lab demonstrated that scientists, after receiving training on how to listen to sonifications, were able to classify moles as cancerous or non-cancerous with 90% accuracy, highlighting the high diagnostic potential of sound (Cotturone, 2025). Similarly, empirical studies in human-computer interaction have shown that while the initial addition of an auditory monitoring task to a visual task may briefly harm performance, users quickly adapt, and the addition of sound to visual interfaces can ultimately enhance overall task performance (Walker & Nees, 2011).

Recent evaluations of complex data sonifications confirm these findings. Hultman et al. (2024) evaluated an interactive sonification tool designed to communicate vessel emissions data in the Baltic Sea. Their empirical results showed that participants in an audio-only group were successfully able to identify ship categories and attach meaning to the emissions data without augmented visual feedback. Furthermore, the sonification proved highly effective at eliciting emotional responses, engaging the users on a deeper level than mere visual statistics. In another study, Tayarzadeh (2025) conducted a mixed-method usability study on a tool that sonified urban form metrics. The quantitative results revealed that while visual-only tasks often yielded the highest overall accuracy, the combination of audio and visual modalities improved response times and user confidence. The results also showed that participants with higher baseline musical perception demonstrated superior performance in interpreting the sonified data. Qualitative feedback from the study further indicated that sonification enhanced emotional engagement, spurred curiosity, and facilitated novel insights into the data, particularly in instances where visual representations were overly cluttered or challenging to interpret.

These past evaluations suggest that SoniMet has the potential to enhance the exploration of bibliometric data, if users are given the opportunity to acclimate to auditory data representations. By continuing to refine the mapping strategies, expanding the scope of analyzable data, and testing the tool with domain experts, metrics sonification can mature from a novel concept into a robust, complementary method for research evaluation in the scientometric community.

# Declaration of generative AI in scientific writing

During the preparation of this work the authors used Gemini Pro 3.1 to improve language and readability. After using this tool, the authors reviewed and edited the content as needed and take full responsibility for the content of the publication.

# Acknowledgements

Lutz Bornmann is a member of the Distinguished Reviewers Board of *Scientometrics*.

# Declarations

The authors did not receive support from any organization for the submitted work. The authors have no relevant financial or non-financial interests to disclose.